\documentclass[runningheads]{llncs}

\usepackage[T1]{fontenc}
\usepackage{graphicx}
\usepackage[numbers,sort&compress]{natbib}
\usepackage{amsmath}
\usepackage{pifont}
\usepackage{amssymb}
\usepackage{xspace}
\usepackage{xcolor}
\usepackage{booktabs}
\usepackage{colortbl}
\usepackage[hidelinks]{hyperref}

\newcommand{\BfPara}[1]{{\vspace{0.5ex}\noindent\colorbox{gray!12}{\bf #1.}}\xspace}
\newcommand{\etal}{{\em et al.}\xspace}

\newcommand{\etc}{{\em etc.}\xspace}

\begin{document}

\title{SPILLOVER: Measuring Cyberbullying Norm Propagation on Social Media}

\author{%
Arslan Bisharat\inst{1} \and
Katelyn Skees\inst{2} \and
Mujtaba Nazari\inst{1} \and
Ayaan Khan\inst{1} \and
Manny Sandoval\inst{1} \and
Mohammed Abuhamad\inst{1} \and
Deborah Hall\inst{2} \and
Yasin Silva\inst{1}%
}

\authorrunning{A. Bisharat et al.}

\institute{%
Loyola University Chicago, Chicago, IL, USA\\
\email{\{marslan,mnazari,akhan95,msandovalmadrigal,mabuhamad,ysilva1\}@luc.edu}
\and
Arizona State University, Glendale, AZ, USA\\
\email{\{kskees,d.hall\}@asu.edu}%
}

\maketitle

\begin{abstract}

While certain aspects of cyberbullying (CB) such as its factors and prevalence have been studied extensively, relatively little attention has been given to specifically how the aggression transfers from comment to comment. This understanding could have important implications for designing better anti-bullying features. In this paper, we study multiple aspects of the nature of this aggression transference in social media sessions.
Using data from 32,754 consecutive comment pairs from 430 Instagram sessions, we find that a preceding CB comment substantially raises the odds of the next comment being CB, an effect confirmed by session fixed-effects controls and driven primarily by cross-user spread.
We also find that $\text{CB} \to \text{CB}$ pairs are more textually similar than $\text{NoCB} \to \text{CB}$ pairs across five complementary methods, and that this pattern holds under a matched cross-session baseline that rules out shared vocabulary, session toxicity, and session length as confounds.
Moreover, non-aggressive replies grow more negative as preceding CB severity increases, a graded pattern consistent with automatic emotional influence below the threshold of overt aggression.
These key findings replicate across three independent datasets (Reddit, Wikipedia Detox, and SOCC), with spillover rates that track platform visibility design.
Finally, we show that a single binary feature (whether the prior comment was CB) improves prediction over session-level baselines and over a fine-tuned HateBERT classifier, serving as a real-time moderation signal that targets the spreading chain rather than individual offenders.
\keywords{Cyberbullying \and Norm propagation \and Cyberbullying spillover \and Social media \and Cross-platform analysis}

\end{abstract}

\section{Introduction}
\label{sec:introduction}

Cyberbullying (CB) is targeted, repeated online aggression toward a specific individual with an underlying power imbalance.
CB-identified comments are known to cluster within conversations, suggesting that cyberbullying spills over into subsequent comments within a post's comments.
Previous work by Kramer \etal\citep{kramer2014experimental} showed that, at the platform level, negative content on Facebook shifts subsequent emotional expression and Cheng \etal\citep{cheng2017anyone} showed that, at the user level, prior negative mood and conversational context predict trolling beyond a user's intrinsic propensity. However, neither study considered how aggression transfers from comment-to-comment within the same session, which is the gap that this paper addresses.

The contributions of this paper are:\footnote{Dataset and code: \url{https://github.com/ysilva/spillover}} \ding{182} the first evidence, to our knowledge, that CB spillover at the level of comment pairs is content-specific rather than generically disinhibiting, with convergent support from five similarity methods and a matched cross-session baseline that controls for shared vocabulary, session toxicity, session length, and topic; \ding{183} a dose-response gradient linking preceding CB severity to sentiment shift in non-CB replies, consistent with emotional contagion; \ding{184} evidence that most cross-user $\text{CB} \to \text{CB}$ commenters are aggressors rather than victims, which separates norm adoption from retaliation; \ding{185} a chain-length decay curve showing the spillover signal persists two to three steps beyond the triggering comment; \ding{186} cross-platform replication across four datasets with spillover rates that vary systematically with platform visibility design; and \ding{187} a real-time binary predictor that improves AUPRC over session-level baselines and over a standalone text classifier, and serves as a directly deployable moderation signal.

\section{Background and Related Work}
\label{sec:background}

The field of cyberbullying detection has evolved considerably, from binary classifiers built on lexical features \citep{dinakar2011modeling,xu2012learning} and deep learning models \citep{agrawal2018deep,caselli2020hatebert} to severity measurement through machine-learning classifiers \citep{PLOSOneSeverity} and the use of victim psychological attributes in detection \citep{prama2025ai}, as well as identity-targeted variants such as LGBTQ+ cyberbullying detection \citep{arslan2024detectinglgbtqinstancescyberbullying} and context-aware attention models \citep{bisharat2026spectrumnet}. Most of these models score each comment independently, with few examining whether conversational context shapes what kind of aggression appears next. Yet how cyberbullying spills over and through what mechanisms remains underexplored, given evidence that aggression spreads online \citep{kramer2014experimental,cheng2017anyone}. Two psychological mechanisms, in particular, predict how a CB comment may transmit an aggressive norm within a session. \textit{Emotional contagion} \citep{hatfield1994emotional} holds that exposure to negative content triggers automatic negative responses, especially when reactions are publicly visible, and research confirms this mechanism extends to online networks \citep{rosenbusch2019multilevel}. The theory predicts that the emotional tone of a conversation should shift after CB exposure and that the magnitude of the shift should scale with the intensity of the exposure. \textit{Moral disengagement} \citep{bandura1999moral,bussey2015moral,paciello2020role} predicts that aggressive behavior by one user signals its acceptability to others, allowing others to displace responsibility for and minimize the perceived consequences of their own aggressive responses. Indeed, online moral disengagement is directly associated with cyberbullying perpetration. This theory predicts not just that other users will become aggressive but that they will adopt the specific content and style of the preceding aggression. Together, these two mechanisms predict a feedback cycle: contagion spreads negative affect while disengagement lowers inhibition and provides a behavioral template.

\section{Analysis and Results}
\label{sec:analysis}
\subsection{Dataset and Pair Construction}
Our primary data source is an Instagram dataset \citep{hosseinmardi2015detection}, extended with comment-level CB severity and topic annotations contributed by our group. The dataset and code are available at \url{https://github.com/ysilva/spillover}. Each Instagram session consists of a post (an image and its accompanying caption) together with all comments it received. The corpus contains 1,559 sessions comprising 106,618 comments. Each comment was annotated by five independent annotators on a four-level severity scale (0 = not CB; 1 = mild CB; 2 = moderate CB; 3 = severe CB), with final labels assigned by majority agreement. A comment is classified as CB if its severity is $\geq 1$, and as NoCB otherwise. The CB rate of a set of comments is the proportion with CB status. The dataset also includes ten topic labels and seven role labels (bully, bully assistant, non-aggressive victim, aggressive victim, non-aggressive defender of the victim, aggressive defender of the victim, and passive bystander). Of the 1,559 sessions, we consider the 430 sessions that contain at least one cyberbullying comment. This restriction is intentional: sessions with no CB comments contribute no CB-initiating events and do not bear on the spreading mechanism. The measured spillover rate is, therefore, scoped to sessions where a spreading chain can activate. Within these 430 sessions, we construct consecutive comment pairs $(c_{i-1}, c_i)$ ordered by timestamp. Each pair is characterized by three properties: the CB status of each comment, the individual severity score (0--3) of each comment, and the time gap between them. Pairs are categorized by the CB status of both comments: $\text{CB} \to \text{CB}$, $\text{CB} \to \text{NoCB}$, $\text{NoCB} \to \text{CB}$, and $\text{NoCB} \to \text{NoCB}$. This process produces 32,754 pairs, which serve as the primary unit of analysis. The pair design isolates the immediate sequential effect of one comment on the next, the most direct test of comment-to-comment contagion. Each claim is, therefore, scoped to immediate sequential spreading rather than long-range dynamics. Because session fixed-effects regression requires a separate intercept for each session, we restrict that particular analysis to sessions with $\geq$20 comments, where enough pairs exist to produce stable estimates. \autoref{tab:spillover_main} summarizes the main dataset statistics. For the cross-platform analysis, we use three additional datasets where a session corresponds to a single submission thread on Reddit, a single talk-page discussion on Wikipedia Detox, and a single article comment section on SOCC. On Reddit, pairs are extracted within each thread in reply-chain order to preserve logical sequential relationships rather than global timestamp order.

\subsection{Occurrence-Level Spillover}
To the extent that cyberbullying spreads, a comment that follows a CB comment should be more likely to be CB than one that follows a noCB comment. We measure this as the simple conditional difference:
\begin{equation}
\Delta_{\text{spillover}} = P(\text{CB}_i \mid \text{CB}_{i-1}) - P(\text{CB}_i \mid \neg\text{CB}_{i-1})
\label{eq:spillover}
\end{equation}
with significance assessed with chi-squared tests and effect sizes reported as Cohen's $d$. Within the 430 CB-containing sessions, the CB rate in $\text{CB} \to \text{CB/NoCB}$ pairs was 49.3\%, compared to 20.5\% in $\text{NoCB} \to \text{CB/NoCB}$ pairs, a 28.8 percentage point gap (\autoref{tab:spillover_main}). The mean severity was also higher in the $\text{CB} \to \text{CB}$ condition, though this effect is small by conventional standards ($d=0.212$). It is possible, however, that certain sessions simply attract more aggressive users and, therefore, produce more consecutive CB comments as a byproduct. In other words, the raw difference could be an artifact of session-level toxicity rather than evidence of a real spreading process. To pull apart within-session influence from between-session differences, we apply three statistical controls. The results of these analyses are presented in the next sub-sections.

\begin{table}[t]
\centering
\small
\setlength{\tabcolsep}{8pt}
\begin{tabular}{@{}lrcc@{}}
\toprule
\textbf{Type of Pair} & \textbf{Pairs ($N$)} & \textbf{CB Rate} & \textbf{Mean Severity} \\
\midrule
\rowcolor{gray!10}
$\text{CB} \to \text{CB/NoCB}$     & 10,877 & 49.3\% & 0.512 \\
$\text{NoCB} \to \text{CB/NoCB}$ & 21,877 & 20.5\% & 0.338 \\
\midrule
\rowcolor{gray!10}
\textbf{Difference} &  & \textbf{28.8\,pp} & \textbf{0.174} \\
\bottomrule
\end{tabular}
\caption{Core properties of the Instagram dataset including number of pairs, CB rates, and mean severity across 32,754 consecutive comment pairs. $N$ = number of pairs. $\text{CB} \to \text{CB/NoCB}$ denotes pairs where the preceding comment is CB; $\text{NoCB} \to \text{CB/NoCB}$ denotes pairs where the preceding comment is not CB. CB rate is 2.4$\times$ higher when the prior comment is CB ($\chi^2{=}2847.3$, $p{<}10^{-275}$). pp = percentage points.}
\label{tab:spillover_main}
\end{table}

\BfPara{Within-Session Validation}
The raw spillover number conflates two things: comment-to-comment association and the background toxicity level of the session as a whole. Consider a session about a controversial political figure: the higher likelihood of toxic language throughout may make consecutive CB comments appear linked even if neither one actually caused the other. To hold session-level factors including a session's baseline toxicity constant, we apply two statistical control strategies. First, a \textit{stratified analysis} divides sessions into five groups by their overall CB rate and recomputes the spillover within each group. If the effect survives even among low-toxicity sessions, it cannot be explained by session composition alone. Second, a \textit{session fixed-effects logistic regression} provides the most rigorous test:
\begin{equation}
\text{logit}\big(P(\text{CB}_i)\big) = \alpha_s + \beta \cdot \mathbf{1}[\text{CB}_{i-1}]
\label{eq:fixed_effects}
\end{equation}
\begin{table}[t]
\centering
\small
\setlength{\tabcolsep}{6pt}
\begin{tabular}{@{}lccc@{}}
\toprule
\textbf{Comment Source} & \textbf{CB $\to$ CB} & \textbf{All post-CB} & \textbf{Per-pair CB Rate} \\
\midrule
\rowcolor{gray!10}
Different user & 82.9\% & 90.3\% & 31.2\% \\
Same user      & 17.1\% &  9.7\% & 59.9\% \\
\bottomrule
\end{tabular}
\caption{Identity tracking of $\text{CB} \to \text{CB}$ pairs by comment source. \textbf{CB $\to$ CB}: fraction of $\text{CB} \to \text{CB}$ pairs from each source. \textbf{All post-CB}: fraction of all pairs where the preceding comment is CB (regardless of the following comment's CB status) from each source. \textbf{Per-pair CB Rate}: the rate at which the following comment is CB, broken down by source. Cross-user spread accounts for 83\% of $\text{CB} \to \text{CB}$ pairs by volume, though same-user pairs show a higher per-pair CB rate (59.9\% vs.\ 31.2\%).}
\label{tab:identity}
\vspace{-5mm}
\end{table}

The session intercept $\alpha_s$ absorbs all between-session variation (topic, audience, and overall toxicity level), so the coefficient $\beta$ captures only within-session influence; that is, each session serves as its own control group. All effects are validated with permutation tests (1,000 iterations) and bootstrap confidence intervals (10,000 resamples). The stratified analysis found positive spillover in all five toxicity quintiles (7.4pp--17.7pp across Q1--Q5). Session heterogeneity is, therefore, not the sole explanation. The fixed-effects model estimated OR $=1.83$ ($p<10^{-61}$), with each session as its own control. As a robustness check, a lower threshold of $\geq$10 comments per session gives OR $=1.79$ ($p<10^{-58}$). The result is, therefore, not specific to longer sessions. To test whether the effect extends beyond the immediate next comment, we defined $k$-step pairs $(c_i, c_{i+k})$ for $k \in \{1,2,3\}$ where $c_i$ is a CB comment, and computed the conditional CB rate of $c_{i+k}$ in each case.
The rate decays from 49.3\% at $k=1$ to 35.1\% at $k=2$ and 26.8\% at $k=3$, compared to the noCB baseline of 20.5\%. This traces a spreading chain that attenuates rather than immediately extinguishes.

\BfPara{Cross-User Spread versus Persistence}
 An alternate explanation to rule out is that a user who posts one CB comment may simply post another, with $\text{CB} \to \text{CB}$ pairs more accurately reflecting individual persistence than a norm that extends beyond a single user. To test this, we track user identity across consecutive pairs and measure what fraction of $\text{CB} \to \text{CB}$ comments come from a different user. Of the CB comments that followed another CB comment, 82.9\% (95\% CI: 81.4\%--84.3\%) were written by a different user (\autoref{tab:identity}). Same-user pairs do show higher per-pair CB rates but cross-user spread accounts for the vast majority of cases by volume. To address whether this cross-user spread reflects victims retaliating rather than new aggressors joining, we examine the role labels of the second commenter in cross-user $\text{CB} \to \text{CB}$ pairs. Only 3.2\% of second commenters are annotated as victims; 84.4\% are bullies or bully assistants. The cross-user effect, therefore, predominantly reflects new aggressors joining the aggressive exchange, not victims responding in kind. This pattern aligns with a moral disengagement account \citep{bandura1999moral} by suggesting that a CB comment lowers the aggression threshold for \textit{other} participants rather than just the original poster. Permutation tests confirmed that the observed spillover exceeds chance ($p<0.001$) and the effect is stable across topics and session lengths. Disaggregating by CB topic using the same conditional OR formula (\autoref{eq:spillover}) applied within each of the ten topic strata reveals heterogeneity in susceptibility: appearance-based attacks show the highest within-session spillover (OR $= 2.1$), followed by racial and ethnic attacks (OR $= 1.9$) and political CB (OR $= 1.7$). This variation across topics is consistent with the moral disengagement account that norm-laden contexts lower the aggression threshold more sharply for bystanders.

\subsection{Sentiment Shift and Dose-Response}
\begin{figure}[t]
\centering
\includegraphics[width=0.69\columnwidth]{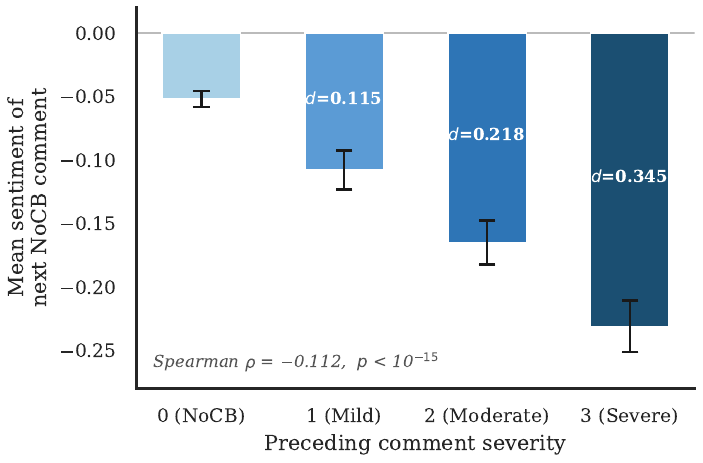}
\caption{Dose-response gradient: mean sentiment of NoCB replies as a function of preceding comment severity. Sentiment grows monotonically more negative; the effect size between severity 0 and severity 3 is $d=0.345$.}
\label{fig:dose_response}
\end{figure}

If the spreading is a genuine contagion process rather than generic disinhibition, then exposure to CB should affect the \textit{emotional tone} of subsequent comments, even those that are not, themselves, aggressive. We score sentiment with VADER and RoBERTa-base-sentiment. The sentiment shift is defined as:

\begin{equation}
\Delta_{\text{sentiment}} = \mathbb{E}[\text{sent}(c_i) \mid \text{CB}_{i-1}] - \mathbb{E}[\text{sent}(c_i) \mid \neg\text{CB}_{i-1}]
\label{eq:sentiment}
\end{equation}
This is restricted to NoCB replies $c_i$, so it captures sub-threshold emotional leakage rather than overt aggression. The dose-response test then checks whether the shift scales with the severity of the preceding comment ($s \in \{0,1,2,3\}$). This distinction is critical: an artifact driven by topic or user selection would produce a uniform increase in negativity, not a graded one. Indeed, noCB comments that followed a CB comment were detectably more negative ($d=0.115$, $p<10^{-13}$; a small effect by conventional standards), a finding confirmed by both RoBERTa and VADER. The dose-response gradient (\autoref{fig:dose_response}) shows that sentiment grows monotonically more negative across all four severity levels (0 through 3); the effect size between the lowest and highest severity levels is $d=0.345$. Given that a confound driven by shared context would produce a uniform shift rather than a graded one, this pattern is consistent with a genuine spillover effect rather than a shared-context artifact.

\begin{table}[t]
\centering
\small
\setlength{\tabcolsep}{4pt}
\begin{tabular}{@{}lccccccc@{}}
\toprule
\textbf{Method} & \textbf{CB$\to$CB} & \textbf{NoCB$\to$NoCB} & \textbf{NoCB$\to$CB} & $d_1$ & $d_2$ & \textbf{XU\,$d$} \\
\midrule
\rowcolor{gray!10}
TF-IDF cosine & 0.112 & 0.060 & 0.044 & 0.418 & 0.276 & 0.328 \\
Word Jaccard  & 0.085 & 0.049 & 0.030 & 0.472 & 0.242 & 0.450 \\
\rowcolor{gray!10}
Char 3-gram   & 0.098 & 0.053 & 0.039 & 0.508 & 0.305 & 0.506 \\
MiniLM-L6-v2  & 0.335 & 0.249 & 0.224 & 0.655 & 0.449 & 0.588 \\
\rowcolor{gray!10}
MPNet-base-v2 & 0.357 & 0.253 & 0.232 & 0.707 & 0.533 & 0.641 \\
\bottomrule
\end{tabular}
\caption{Mean pairwise text similarity for consecutive comment pairs across five methods. NoCB$\to$NoCB is used as a general conversational coherence baseline. $d_1$: Cohen's\,$d$ for CB$\to$CB vs NoCB$\to$CB; $d_2$: Cohen's\,$d$ for CB$\to$CB vs NoCB$\to$NoCB; XU\,$d$: cross-user pairs only ($d_1$). NoCB$\to$NoCB pairs are slightly more similar than NoCB$\to$CB pairs, confirming a small general conversational coherence effect. CB$\to$CB similarity exceeds this baseline by a factor of $2$--$3$. All $p{<}10^{-100}$.}
\label{tab:content_similarity}
\end{table}

\subsection{Content-Level Spreading} Because emotional tone shifts on their own do not prove that specific \textit{content} transfers, a more stringent test is whether the topics and language of CB carry over from one comment to the next. That is, if cyberbullying spillover is content-specific, then $\text{CB} \to \text{CB}$ pairs should be more linguistically similar than $\text{NoCB} \to \text{CB}$ pairs. We study content transfer using five methods spanning lexical and semantic representations. The three lexical methods (TF-IDF cosine, word Jaccard, character 3-gram) share surface-level vocabulary signal and are treated as one convergent line of evidence. The two neural sentence encoders (MiniLM-L6-v2 \citep{reimers2019sentence} and MPNet-base-v2 \citep{song2020mpnet}) capture semantic similarity independently of exact word overlap and constitute a second, independent line of evidence. We compare text similarity between $\text{CB} \to \text{CB}$ and $\text{NoCB} \to \text{CB}$ pairs. Both pair types are drawn from the same sessions, so any shared session vocabulary affects both equally. A content-transfer process would predict a consistent $\text{CB} \to \text{CB}$ advantage. All five text similarity methods confirmed that $\text{CB} \to \text{CB}$ pairs are more textually similar than $\text{NoCB} \to \text{CB}$ pairs (\autoref{tab:content_similarity}), with effect sizes from $d=0.42$ to $d=0.71$. These effects persist after removing same-author pairs ($d=0.33$--$0.64$ for cross-user only). To assess whether the elevated $\text{CB} \to \text{CB}$ similarity reflects something specific to cyberbullying or merely general conversational coherence, we include $\text{NoCB} \to \text{NoCB}$ pairs as a third reference condition. $\text{NoCB} \to \text{NoCB}$ pairs are slightly more similar than $\text{NoCB} \to \text{CB}$ pairs ($d=0.11$--$0.17$). This confirms a small background coherence effect: conversations, in general, tend to stay somewhat on-topic. However, $\text{CB} \to \text{CB}$ similarity substantially exceeds this background level ($d=0.24$--$0.53$ across all five methods; \autoref{tab:content_similarity}), an excess $2$--$3{\times}$ larger than the general same-type coherence. The within-type similarity advantage is not simply a property of consecutive comments sharing a topic; it is markedly stronger for cyberbullying transitions. Users are not just becoming generically aggressive after exposure to CB: they adopt the specific language and style of the preceding aggression to a degree that exceeds ordinary conversational drift. Together with the dose-response gradient, this pattern is consistent with content-specific contagion rather than generic disinhibition.

\BfPara{Ruling Out Vocabulary Confounds}
A natural objection to the content similarity finding is a shared-label vocabulary confound: CB comments may appear similar to each other simply because both are CB and, therefore, share the aggressive vocabulary common to cyberbullying, not because one caused the other. We construct a matched cross-session baseline to test this directly. For each within-session $\text{CB} \to \text{CB}$ pair, we retain the preceding comment but replace the following CB comment with a randomly-sampled CB comment drawn from a \textit{different} session. Both the within-session and cross-session groups, therefore, contain two CB comments. To ensure a fair comparison, we stratify sessions into five quintile bins on CB rate and five quintile bins on session length to form 25 match cells, then sample replacement comments from the same cell. We run 1,000 independent bootstrap samples and average scores across samples for stability. This produces four comparison conditions: unmatched, toxicity-matched, fully-matched (toxicity and length combined), and the standard $\text{NoCB} \to \text{CB}$ baseline. Within-session $\text{CB} \to \text{CB}$ pairs remain substantially more similar than the fully-matched cross-session baseline across all five methods ($d = 0.51$--$1.03$, all $p \approx 0$). The marginal contribution of each matching dimension is small: adding toxicity matching shifts the baseline by $d = 0.04$--$0.13$ and adding length matching adds a further $d = 0.04$--$0.07$. This means that even the unmatched cross-session comparison already controlled for the bulk of compositional differences; the main effect is robust to both confounds (CB rate and session length). To test whether shared topic labels drive any residual gap, we additionally restrict to pairs that share at least one topic label and apply the same matching procedure within each topic. This topic-stratified ablation produces $d = 0.61$--$1.13$, larger than the unrestricted estimate because it isolates pairs where topic alignment is guaranteed on both sides. The consistent within-session advantage across all conditions makes shared aggressive vocabulary, session toxicity, length, or topic unlikely as the primary drivers of the elevated similarity, though observational data cannot fully rule out unobserved confounders.

\subsection{Spreading-Aware Prediction}
If the spreading signal is real, it should improve the ability to anticipate CB before it occurs. We frame this as a binary classification task: can we predict whether $c_i$ is CB \textit{without access to $c_i$'s text}? We test three conditions. The \textit{sole-signal} condition uses a single binary feature: was the prior comment CB? The \textit{baseline} model uses 7 session-level features (position, session length, time elapsed, \etc). The \textit{CB-signal} model adds the binary CB-status feature to the baseline. Because the positive class is rare (9.2\%), we report Average Precision (AUPRC) as the primary metric: a random classifier would attain AUPRC equal to the prevalence (0.092). Gains are, therefore, directly interpretable as improvement over chance. All conditions are tested on Logistic Regression, Random Forest, and Gradient Boosting with 5-fold session-grouped cross-validation. Improvements are assessed with paired $t$-tests on fold-level AUPRC. Used alone, the prior comment's CB status attains AUPRC\,$=0.289$ (a 3.2$\times$ lift above the random baseline of 0.092) and ROC-AUC\,$=0.72$. This confirms the spillover signal carries genuine predictive information even in isolation. When added to the session-level baseline, this single binary feature improves AUPRC by $+0.031$--$+0.037$ ($+6.6$--$8.2$\%; $p<0.01$ across all three classifiers). A per-feature ablation across five candidate predecessor features confirms that \textit{prev\_is\_cb} is the only feature that is positive and statistically significant across all three classifiers; timing, authorship, and text-length features are inconsistent or near zero (\autoref{tab:prediction}). The spillover signal should be used as a complement to text-based classifiers, not a replacement. Practically, the use of this signal could be helpful for early detection of cyberbullying and proactive moderation. To verify the signal is orthogonal to text-level evidence, we fine-tuned a HateBERT \citep{caselli2020hatebert} classifier on the same 430-session folds (3 epochs, learning rate $2{\times}10^{-5}$, batch size 16; \textit{prev\_is\_cb} was appended as a binary feature to the pooled [CLS] representation before the final classification layer). As \autoref{tab:hatebert} shows, HateBERT alone attains AUPRC $= 0.71$; adding \textit{prev\_is\_cb} as a single additional feature raises this to $0.74$ ($+0.03$, $p<0.01$). The spillover signal, therefore, captures information orthogonal to comment text.
\begin{table}[t]
\centering
\small
\setlength{\tabcolsep}{4.5pt}
\begin{tabular}{@{}llccccc@{}}
\toprule
\textbf{Model} & \textbf{Met.} & \textbf{Baseline} & \textbf{+CB Status} & \textbf{$\Delta_1$} & \textbf{+Content} & \textbf{$\Delta_2$} \\
\midrule
\rowcolor{gray!10}
Logistic Regression  & AUC & 0.713 & 0.740 & $+$0.027$^{***}$ & 0.740 & $<$0.001\textsuperscript{n.s.} \\
Random Forest        & AUC & 0.705 & 0.732 & $+$0.027$^{***}$ & 0.740 & $+$0.008$^{***}$ \\
\rowcolor{gray!10}
Gradient Boosting    & AUC & 0.684 & 0.716 & $+$0.032$^{***}$ & 0.727 & $+$0.012\textsuperscript{n.s.} \\
Gradient Boosting    & F1  & 0.292 & 0.339 & $+$0.047$^{***}$ & 0.374 & $+$0.035$^{**}$  \\
\bottomrule
\end{tabular}
\vspace{2pt}
\caption{Spillover-aware CB prediction: three-condition ablation across classifiers. Adding the CB status of the prior comment (+CB Status) is the primary driver of improvement; content features (+Content) add a small further increment. $\Delta_1$ = gain over Baseline; $\Delta_2$ = gain over +CB Status. $^{***}p<0.001$; $^{**}p<0.01$; n.s.\,=\,not significant. Paired $t$-tests on 5-fold AUC/F1 scores.}
\label{tab:prediction}
\end{table}

\begin{table}[t]
\centering
\small
\setlength{\tabcolsep}{5pt}
\begin{tabular}{@{}lcc@{}}
\toprule
\textbf{Model} & \textbf{AUPRC} & \textbf{$\Delta$} \\
\midrule
\rowcolor{gray!10}
Sole signal (\textit{prev\_is\_cb} only)  & 0.289 & \multicolumn{1}{c}{--} \\
HateBERT (text only)                       & 0.710 & \multicolumn{1}{c}{--} \\
\rowcolor{gray!10}
HateBERT + \textit{prev\_is\_cb}           & 0.740 & $+$0.030$^{**}$ \\
\bottomrule
\end{tabular}
\vspace{2pt}
\caption{HateBERT integration: adding \textit{prev\_is\_cb} to a fine-tuned HateBERT classifier. The spillover signal captures information orthogonal to comment text.}
\label{tab:hatebert}
\end{table}

\subsection{Cross-Platform Replication} The results discussed thus far came from Instagram, whose flat comment layout and younger users may amplify the effect. Our final analysis tested the generalizability of the key findings across three additional platforms with different interaction designs. Specifically, we performed the core spillover, content similarity, and identity analyses on Reddit \citep{baumgartner2020pushshift} (threaded comments from five subreddits spanning news, politics, and community discussion; 2015--2019; 314,588 pairs), Wikipedia Detox \citep{wulczyn2017ex} (English page discussions; 2004--2015; 63,155 pairs) and SOCC \citep{kolhatkar2020socc} (Globe \& Mail news comments; 2012--2016; 338,909 pairs). Because these datasets lack human severity annotations, the toxicity labels are generated with the Detoxify unbiased classifier \citep{Detoxify} at a threshold of 0.5. Detoxify's toxicity construct is broader than cyberbullying, which requires targeted harm and repetition; the cross-platform findings, therefore, reflect toxicity spillover and should be interpreted with that scope in mind. To validate label consistency, we applied Detoxify to the Instagram data and compared its predictions against human annotations. The comparison gives AUC $= 0.91$, which supports treating the automated labels as a reasonable proxy for the cross-platform analyses. The cross-platform results are also robust to threshold choice: repeating all analyses at thresholds of 0.3 and 0.7 gives spillover ORs within 0.08 of the values reported at 0.5 across all three platforms. This rules out threshold sensitivity as an explanation for the observed effects. The spillover effect replicated on all three platforms (all $p<10^{-61}$, \autoref{tab:cross_platform}). Wikipedia Detox showed the strongest lift (1.95$\times$), followed by SOCC (1.42$\times$) and Reddit (1.26$\times$). This ordering effect tracks platform design: Wikipedia talk pages and SOCC comment sections display comments in a linear layout where prior content stays visible and susceptibility to contagion is high. Reddit's nested threading distributes replies across sub-threads and dilutes the sequential signal. The lift's systematic variation with users' prior content exposure supports a visibility-driven contagion mechanism. Content-level spreading also replicated on all platforms (\autoref{tab:cross_platform}) and cross-user spread was the primary driver on every platform (90--100\%; the 100\% figure for Wikipedia Detox reflects the structural constraint that Wikipedia edits are always by different users).

\begin{table}[t]
\centering
\small
\setlength{\tabcolsep}{2pt}
\begin{tabular}{@{}lrrcccr@{}}
\toprule
\textbf{Platform} & \textbf{Pairs} & \textbf{Sessions} & \textbf{Lift} & \textbf{Content $d$} & \textbf{Cross-User} & \textbf{$p$-value} \\
\midrule
\rowcolor{gray!10}
InstaCTSR & 32,754 & 430 & 1.83$\times$ & 0.42--0.71 & 82.9\% & $<10^{-61}$ \\
Wikipedia Detox        & 63,155 & 15,789 & 1.95$\times$ & 0.537 & 100.0\% & $<10^{-275}$ \\
\rowcolor{gray!10}
SOCC (Globe \& Mail)   & 338,909 & 3,703 & 1.42$\times$ & 0.362 & 93.8\% & $<10^{-61}$ \\
Reddit (5 subreddits)  & 314,588 & 2,440 & 1.26$\times$ & 0.264 & 97.6\% & $<10^{-124}$ \\
\bottomrule
\end{tabular}
\caption{Cross-platform replication of the spillover effect. Lift = ratio of CB rate after CB to CB rate after NoCB. Content $d$ = Cohen's $d$ for text similarity between CB$\to$CB and NoCB$\to$CB pairs (word Jaccard for Reddit, overall for others). Cross-User = percentage of CB$\to$CB pairs from different users.}
\label{tab:cross_platform}
\end{table}

\section{Discussion}
\label{sec:discussion}

\BfPara{Not Just More, But Alike}
Cyberbullying clustering in online conversations is well known. This paper provides new evidence of immediate comment-to-comment spillover within conversations; our content-level analyses suggest that this spillover is not a session-level artifact. Whereas prior work has shown that exposure to aggression makes more aggression likely \citep{cheng2017anyone,kramer2014experimental}, our findings suggest that it makes \textit{the same kind} of aggression likely. $\text{CB} \to \text{CB}$ pairs are more textually similar than $\text{NoCB} \to \text{CB}$ pairs across five methods spanning lexical and semantic representations and this holds after the removal of same-author pairs. A shared situational trigger would elevate CB rates without the texts becoming more alike, but the texts \textit{are} more alike. Critically, including $\text{NoCB} \to \text{NoCB}$ pairs as a baseline reveals that CB-specific within-type similarity ($d=0.24$--$0.53$ above the $\text{NoCB} \to \text{NoCB}$ level) is $2$--$3{\times}$ larger than the general conversational coherence effect ($d=0.11$--$0.17$). This rules out the possibility that the finding merely reflects topic drift common to all same-type pairs. This illustrates an important distinction between generic disinhibition (whereby exposure to aggression increases subsequent aggression, broadly) and content-specific spillover (where, for example, exposure to a religion-based attack increases the likelihood of a subsequent religion-based attack). This distinction has direct consequences for cyberbullying classifiers. Most severity classifiers score comments independently of conversational context, capturing a snapshot of the text but not the conversational process that shaped it. If CB spillover shapes the language of subsequent comments, then severity labels may partly reflect conversational history rather than the text alone, a hypothesis testable by presenting the same comment to annotators in different conversational contexts. Our HateBERT experiment shows that preceding CB status is not fully captured by the comment text, so incorporating it as a contextual feature offers a direct model improvement.

\BfPara{Interpreting the Evidence}
Whereas observational data cannot conclusively establish causation, our findings offer evidence that goes beyond mere association. The common-cause alternative, that unobserved time-varying confounders independently and chiefly effect consecutive comments, is difficult to reconcile with three features of the data. First, the sentiment shift scales monotonically with preceding severity (the dose-response gradient). A common cause would elevate aggression uniformly regardless of the preceding comment's severity level, not produce a graded pattern. This graded pattern is more consistent with genuine spillover than with a session-level artifact. Second, the content similarity results show that $\text{CB} \to \text{CB}$ pairs are not just more frequent but more textually similar. $\text{CB} \to \text{CB}$ similarity exceeds the background conversational coherence established by $\text{NoCB} \to \text{NoCB}$ pairs by a factor of $2$--$3$, a gap that a common-cause account driven by shared topic or session context alone would not predict. Third, the matched cross-session baseline confirms that the similarity advantage cannot be attributed to shared aggressive vocabulary, session toxicity, session length, or topic alignment. Within-session pairs remain substantially more similar than the matched cross-session comparison ($d = 0.51$--$1.03$, all $p \approx 0$), and the topic-stratified ablation strengthens rather than weakens this finding ($d = 0.61$--$1.13$). These three lines of evidence, together with the cross-platform replication, are collectively more consistent with a genuine spillover effect than with common-cause alternatives. The spillover rate also varies with platform design: strongest on linear-comment platforms (Wikipedia: 1.95$\times$, SOCC: 1.42$\times$) and weakest on Reddit (1.26$\times$), where threading dilutes sequential exposure. A statistical artifact would not co-vary systematically with platform design.

\BfPara{Implications}
 The present findings are consistent with both emotional contagion and moral disengagement, though observational data cannot establish these mechanisms conclusively. Emotional contagion \citep{hatfield1994emotional,rosenbusch2019multilevel} predicts that exposure to negativity triggers automatic negative responses that scale with the intensity of exposure. The observed dose-response gradient aligns with this prediction: even noCB comments become more negative after CB exposure, and the effect tracks severity. Moral disengagement \citep{bandura1999moral,bussey2015moral,paciello2020role} predicts that modeled aggression provides permission and a behavioral template. Content similarity results are consistent with users adopting specific themes and language of preceding aggression, not just general negativity. The cross-user finding (82.9\% of successive CB from different users) is consistent with moral disengagement, as prior comments may lower the aggression threshold for others.
The prediction analysis shows that there is a usable signal in the conversational context. A single binary feature (whether the prior comment was CB) attains AUPRC\,=\,0.289 in isolation (3.2$\times$ above chance) and improves AUPRC by $+0.031$--$+0.037$ when added to a session-level baseline ($p<0.01$ across all classifiers). This opens the door to conversation-level interventions. At a threshold of 0.40, the sole-signal model attains precision 0.38 and recall 0.51 on held-out folds; approximately 62\% of flagged comments are false positives, so \textit{prev\_is\_cb} is best used as a priority signal to focus reviewer attention rather than as an automatic block. One further caveat for deployment is adversarial adaptation: if users learn that inserting a noCB comment resets the signal, they may do so to break the detection chain, a standard arms-race concern for any rule-based moderation trigger. Content downranking could break the transmission chain by reducing the visibility of a comment. Warning labels could counteract moral disengagement by making other users more explicitly aware of aggressive content and an emerging pattern within a session. All of these target the spreading mechanism rather than any single offender. Notably, within-session analyses show that the spillover effect does not decay with the time gap between comments. This pattern is more consistent with content visibility than with emotional urgency as the primary driver, though both may operate simultaneously. Content-level interventions such as warnings, downranking, and flagging are, therefore, likely to be more actionable than timing-based constraints alone.

\BfPara{Limitations} \label{sec:limitations} The primary Instagram dataset was collected in 2012--2014; platform interfaces and moderation policies have since evolved. While cross-platform replication (Reddit, Wikipedia Detox, and SOCC) supports generalizability, these external datasets use Detoxify automated labels rather than human annotation. Detoxify attains $\geq$0.95 AUC on its test set and the directional findings replicate those from human-annotated Instagram data. The unit of analysis in this work is the consecutive comment pair, capturing immediate sequential influence but not longer-range trajectories. We extend to three-step chains, but graph-based analyses could reveal dynamics beyond this window. Two confounds bear on the content similarity results. For shared aggressive vocabulary, the matched cross-session baseline addresses this directly: within-session pairs remain more similar than matched cross-session pairs ($d = 0.51$--$1.03$) and the topic-stratified ablation ($d = 0.61$--$1.13$) rules out shared topic labels. For temporal escalation, the dose-response gradient scales with the severity of the immediately preceding comment rather than session stage, and the 82.9\% cross-user spread indicates the effect holds across different participants. Experimental validation remains the definitive next step. A within-session time-varying confound also warrants acknowledgment: a viral post or trending controversy could drive many users to comment aggressively in rapid succession without any transmission between comments. The session fixed-effects model absorbs between-session variation but does not eliminate this within a session. The dose-response gradient and the matched cross-session baseline provide partial evidence against this alternative, but only a controlled experiment can fully rule it out.

\BfPara{Future Work}
Although our cross-platform replication confirms that the spill- over effect generalizes, the observational evidence warrants further investigation to more conclusively establish causation. A natural next step is randomized content-delay interventions with platform partners, which could examine whether breaking the transmission chain actually reduces downstream aggression. Ephemeral platforms like Snapchat and real-time interactions like gaming chat remain untested and may reveal different spreading dynamics, given that message persistence and visibility differ from the comment threads studied here. Finally, user-specific psychological and demographic factors \citep{prama2025ai} could reveal who is most susceptible to content-level contagion.

\section{Conclusion}
\label{sec:conclusion}

Whereas prior work has shown that online toxicity clusters, this paper shows that cyberbullying, a more specific construct than general online toxicity, spills over immediately from one comment to the next within conversations and that this spillover is content-specific. Across 32,754 comment pairs from 430 Instagram sessions, cyberbullying occurrence spreads within conversations (OR $=1.83$, $p<10^{-61}$) and the effect is primarily cross-user (82.9\% of successive CB comes from different users). Content-level analyses provide evidence consistent with genuine content-specific spillover. $\text{CB} \to \text{CB}$ pairs are more textually similar than $\text{NoCB} \to \text{CB}$ pairs across five similarity methods spanning lexical and semantic representations ($d=0.42$--$0.71$), sentiment in non-aggressive replies shifts in proportion to preceding severity ($d=0.345$), and these patterns replicate across Reddit, Wikipedia Detox, and SOCC with  spillover rates that track platform design (1.26$\times$ to 1.95$\times$). A single binary feature (whether the prior comment was CB) attains AUPRC\,=\,0.289 as the sole predictor (3.2$\times$ above the random baseline of 0.092), improves AUPRC by $+0.031$--$+0.037$ when added to a session-level baseline ($p<0.01$ across three classifiers), and raises a fine-tuned HateBERT classifier from 0.71 to 0.74 ($+0.03$, $p<0.01$), orthogonal to comment text. This translates the spillover effect into a real-time moderation signal for early cyberbullying detection and proactive moderation.

\subsubsection*{Acknowledgements.}
This research was supported in part by National Science Foundation Awards No.~2435164 and No.~2435165, a Google Award for Inclusion Research, and a Lambda Research Grant.

\bibliographystyle{splncs04}
\bibliography{custom}

\end{document}